# GPUArmor: A Hardware-Software Co-design for Efficient and Scalable Memory Safety on GPUs

Mohamed Tarek Ibn Ziad, Sana Damani, Mark Stephenson, Stephen W. Keckler, and Aamer Jaleel
NVIDIA
{mtarek, sdamani, mstephenson, skeckler, ajaleel}@nvidia.com

**Abstract**

Memory safety errors continue to pose a significant threat to current computing systems, and graphics processing units (GPUs) are no exception. A prominent class of memory safety algorithms is *allocation-based* solutions. The key idea is to maintain each allocation's metadata (base address and size) in a disjoint table and retrieve it at runtime to verify memory accesses. While several previous solutions have adopted allocation-based algorithms (e.g., cuCatch [36] and GPUShield [16]), they typically suffer from high memory overheads or scalability problems. In this work, we examine the key characteristics of real-world GPU workloads and observe several differences between GPU and CPU applications regarding memory access patterns, memory footprint, number of live allocations, and active allocation working set. Our observations motivate GPUArmor, a hardware-software co-design framework for memory safety on GPUs. We show that a simple compiler analysis combined with lightweight hardware support using a small Memory Lookaside Buffer (MLB) can help prevent spatial and temporal memory violations on modern GPU workloads with 2.3% average run time overheads. More importantly, GPUArmor achieves speed-of-light performance with negligible storage requirements. This result benefits both base and bounds solutions and memory tagging techniques, which we showcase with GPUArmor-HWOnly, a variation of GPUArmor that does not require recompilation, and achieves 2.2% slowdowns while significantly reducing storage overheads beyond traditional memory tagging approaches.

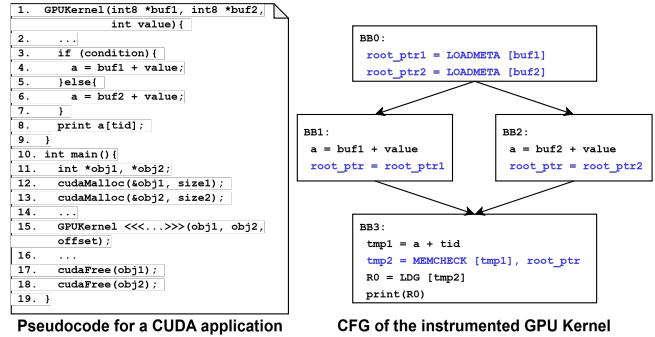

**Figure 1: GPUArmor compiler instrumentation.**

## 1 Introduction

The lack of memory safety in C and C++ is a serious and long-standing issue on CPUs [32, 34, 38]. Modern GPU programming languages, such as CUDA and OpenACC, are extensions of C++ and also lack guarantees regarding the spatial and temporal validity of memory accesses. Consequently, it is unsurprising that recent research has shown that GPU software stacks are susceptible to the same vulnerabilities affecting CPUs for decades [6, 12, 18, 26]. For example, the Mind-Control attack [26] exploits a buffer overflow vulnerability on GPUs to reduce the prediction accuracy of machine learning workloads. Similarly, [12] identifies several memory safety vulnerabilities in modern GPUs, posing significant security risks to GPU applications. Thus, memory safety on GPUs is imperative.

Over the years, researchers and engineers have developed various memory safety solutions to mitigate the issues caused by unsafe programming languages. Techniques that have gained widespread adoption for protecting memory-unsafe languages on CPUs rely on *shadow memory* to store run-time metadata. For instance, the *AddressSanitizer* tool for CPUs associates a byte of shadow memory with every eight bytes of application memory [29]. At run-time, the tool verifies the validity of a pointer dereference by checking the value at the corresponding shadow memory location. AddressSanitizer uses a *direct addressing* hash table, which provides $O(1)$ access time [5]. Unfortunately, the need for speed in AddressSanitizer results in a significant 12.5% (i.e., 1/8) memory bloat.

GPUShield [16] and cuCatch [36], two state-of-the-art GPU memory protection solutions, also use direct addressing tables to facilitate fast verification of memory operations. GPUShield's hash table is manageably sized because the universe of its keys is small, thereby limiting the scalability of the technique [16]. For example, GPUShield cannot handle five of the workloads we studied because they have more than 128 live allocations. cuCatch is scalable, but like AddressSanitizer, it suffers a 12.5% memory bloat [36]. Both schemes use a key (*e.g.*, a pointer identifier [16] or a data address [36]) to retrieve metadata (*i.e.*, the base and bounds) for the allocation to which the key maps in the hash table. The choice of direct addressing tables is presumably based on the conventional wisdom that memory operations are frequent and tend to be on the critical path, so any additional latency a memory protection scheme incurs will directly translate to run time overheads.

This paper presents GPUArmor, a memory protection scheme that challenges the wisdom of using direct addressing tables for GPU memory protection. Initially, we planned for GPUArmor to provide straightforward hardware acceleration for cuCatch, a comprehensive pure software solution for GPU memory protection that delivers base and bounds protection for *global*, *shared*, and *local* memory spaces. We sought to implement hardware instructions for the two intrinsics that cuCatch requires: LOADMETA and MEMCHECK [36]. Notably, LOADMETA is the intrinsic function in cuCatch that fetches the metadata from the hash table for the allocation associated with a data address.

Within well-defined ABI boundaries (i.e., within a function or kernel), cuCatch finds the set of root pointers from which all other pointers within the boundary are derived, and eagerly converts them to fat pointers. Figure 1 illustrates how cuCatch functionally transforms the kernel, GPUKernel, to protect the pointer deference a[tid] on line (8). An analysis determines that buf1 and buf2 are the root pointers for the kernel, and inserts LOADMETA functionality at the kernel's entry to lookup the metadata for buf1 and buf2 from the hash table, effectively turning them into fat pointers. The fat pointer metadata propagates through pointer arithmetic (the blue instructions in blocks BB1 and BB2), which the MEMCHECK function in BB3 uses to verify that a + tid is temporally valid and within the bounds of the root pointer from which 'a' is derived.

Our initial implementation of GPUArmor followed cuCatch's structure. We introduced two new ISA-level instructions for cuCatch's intrinsics along with their microarchitectural implementations. Our results were surprisingly insensitive to LOADMETA's latency. We analyzed real-world GPU workloads to understand our counterintuitive results, and discovered critical differences between CPU and GPU applications that directly impact GPU memory protection schemes. For example, we notice stark differences in memory access patterns, memory footprint, allocation working set size, and the number of live allocations. In particular, we noticed that while the size and number of live allocations tend to be large, the allocation working set is fairly small. These studies motivated a small hardware-managed structure to store recently fetched metadata, which we call the Metadata Lookaside Buffer (MLB). The MLB is extremely effective, and with it, GPUArmor can consider different points in the *time-space tradeoff* of hash tables [28] for LOADMETA's operation, paying for retrieval time primarily for compulsory misses. A hash table with $O(N)$ time retrieval cost (*i.e.*, a linked list of allocation records) incurs negligible space overhead and less than 10% run time overhead, while an $O(logN)$ structure limits run time overhead to under 2%, on par with a speed-of-light solution.

While we base GPUArmor's design on cuCatch, we emphasize that our insights apply to other memory protection schemes with structural similarities. For example, using a variation of GPUArmor, we apply our approach to memory tagging, attaining 2% average run time overhead with insignificant memory bloat.

In this paper we make the following contributions:

- We analyze real-world GPU workloads, uncovering new insights into memory allocation properties.
- We propose and evaluate a hardware-software co-design for scalable and efficient memory safety for GPUs.
- We show that a small cache marginalizes a GPU memory safety scheme's metadata lookup cost, freeing designers to choose slow but space-efficient structures.
- We demonstrate that our approach reduces memory requirements of widely-used memory tagging algorithms.

## 2 Background and Related Work

We now overview memory safety vulnerabilities, describe the GPU programming model with its different memory spaces, and summarize prior memory safety schemes and their limitations. This paper uses the terms *object* and *allocation* interchangeably to refer to a memory allocation.

### 2.1 Memory Safety

Applications written in C, C++, and CUDA, are vulnerable to a variety of memory safety errors because they do not validate the bounds and lifetime of memory accesses. Memory safety errors can lead to control-flow hijacking, silent data corruption, difficult-to-diagnose crashes, and security exploitation [34] even in the presence of trusted execution environments (TEEs) [3].

**Spatial Memory Safety.** This class of memory safety errors occurs when a pointer is used to access an allocation beyond its intended bounds (i.e., base address and size), such as buffer over- or underflows. If the overflow target is adjacent to the victim buffer, it is called linear overflow (e.g., using a large "size" argument in a memcpy call-site). On the other hand, if the overflow target is non-adjacent to the victim buffer, it is referred to as non-linear overflow (e.g., using an arbitrarily large array index, a[index]).

**Temporal Memory Safety.** This class of errors occurs when an application uses a pointer to access an allocation beyond its lifetime, such as use-after-free (UAF). If the application uses a dangling pointer to access a heap allocation after it is deleted, it is referred to as immediate UAF. If the dangling pointer is used after the deleted memory is reallocated, it is called delayed UAF or use-after-realloc.

### 2.2 GPU Memory Spaces

While our work applies broadly to GPUs, we focus on NVIDIA GPUs and the CUDA programming model due to their popularity and importance in safety-critical applications. A CUDA application consists of host-side functions, which run on the CPU, and device-side functions (called kernels), which run on the GPU with thousands of concurrent threads. Device-side GPU kernels can store user data in different memory spaces including: local memory, which is thread-private and cannot be shared, shared memory, which is only shared among threads running on the same streaming multiprocessors (SMs), and global memory (the heap), which is universally shared among all threads running on the GPU. None of our workloads uses local memory, reflecting the reality that local memory allocations are rare. We therefore focus on protecting global and shared memory spaces and argue for reverting to a software-only approach for local memory protection [36].

### 2.3 Memory Safety on GPUs

Existing GPU-based memory safety solutions can be grouped into three categories, as shown in Table 1.

**Tripwires.** This category of GPU solutions distinguishes allocated from unused memory regions by either tracking allocated memory ranges [22] or surrounding them with canaries [7, 8, 10, 11]. Existing tripwire-based solutions suffer from high run time overheads (as they are implemented in software) and are ineffective against non-adjacent out-of-bound (OOB) errors.

**Memory Tagging.** These schemes rely on tag mismatches to probabilistically detect the errors missed by tripwires. A recent implementation of memory tagging on GPUs is LAK [40], which uses a physical memory carve-out for tag storage (8-bit tags per 16B granularity). Carve-out based memory tagging implementations are

Mohamed Tarek Ibn Ziad, Sana Damani, Mark Stephenson, Stephen W. Keckler, and Aamer Jaleel
NVIDIA
{mtarek, sdamani, mstephenson, skeckler, ajaleel}@nvidia.com

Table 1: Comparison with prior memory safety works on GPUs.

| | Proposal | Technology | Spatial Prot. [†] | Temporal Protection | Protection Scalability | Metadata Type | Storage Costs | Run time Overheads [‡] |
|---|---|---|---|---|---|---|---|---|
| Tripwires | Compute Sanitizer [22] | Binary | ◐ | **None** | Yes | Disjoint | 2 words per object | High |
| | GMOD [7, 8] | Compiler | ◐ | **None** | Yes | Adjacent | 8B canary per object | Low |
| | clARMOR [10, 11] | Compiler | ◐ | **None** | Yes | Adjacent | 8B canary per object | Moderate |
| Memory Tagging | LAK [40] | Hardware | ◐ | Probabilistic | Yes | Co-joined | 4 bits per 16B region | Moderate |
| | IMT [33] | Hardware | ◐ | Probabilistic | Yes | Co-joined | Embedded within ECC | None |
| | **GPUArmor-HWOnly** | Hardware | ◐ | Probabilistic | Yes | Co-joined | 2 words per object | Low |
| Base & Bounds | cuCatch [36] | Compiler | ● | Probabilistic | Yes | Co-joined | 32 bits per 32B region | High |
| | GPUShield [16] | Compiler+HW | ● | **None** | **No** | Co-joined | 2 words per object | Low |
| | **GPUArmor** | Compiler+HW | ● | Probabilistic | Yes | Co-joined | 2 words per object | Low |

[†] Spatial Safety Protection: ● - Complete (Adjacent and non-adjacent OOBs); ◐ - Adjacent OOBs only; ○ - No detection.
[‡] Run time overheads based on reported average in original paper: High is > 10%, Low is ≤ 2%, Moderate is 2-10%

storage inefficient. A 32KB object consumes a 2KB metadata storage for maintaining the same 8-bit tag value. Another GPU-based memory tagging proposal is IMT [33]. While IMT completely avoids the storage and run time overheads of performing the memory safety checks by embedding the memory tag bits in ECC, IMT has two main limitations. First, IMT fails to detect tag mismatches in the presence of data errors. While the probability of having a tag mismatch and a random error affecting the same memory word is rare, adversaries can leverage RowHammer [15] to induce errors for bypassing the tag check even in the presence of ECC [4]. Second, IMT is constrained by the underlying ECC properties such as granularity (e.g., 2B of ECC per 32B data access or 6.25% redundancy), which either results in memory fragmentation (due to padding allocations to 32B) or undetected sub-32B overflows.

**Base and Bounds.** The third category of GPU memory safety solutions provides the highest spatial safety coverage. One example is cuCatch, which is a software-only memory safety debugger that opportunistically creates fat pointers with base and bounds protection within ABI boundaries [36]. As a software-only solution, cuCatch suffers from non-negligible run time and memory overheads. Another example is GPUShield, which uses the upper pointer bits to index into a per-kernel bounds table for retrieving the bounds of the pointed-to object [16]. While GPUShield has low run time overheads and deterministic spatial safety guarantees, it has three main limitations. First, GPUShield can only protect a limited set of allocations (bounded by the number of currently unused upper pointer bits). This maximum object count (128 on 57-bit architectures) can be easily exceeded in case of (1) in-kernel `malloc`s, where each thread allocates its own object and (2) HMM allocations that are created on the host-side and implicitly migrated to the GPU. Second, the implicit migration poses another challenge for GPUShield, which only protects explicitly-passed allocations to device-side code via kernel arguments whereas HMM allocations can be passed to device-side code as a pointer within a `struct`, escaping GPUShield's instrumentation. Third, GPUShield creates an immutable metadata table for each kernel upon launch, which persists for the kernel's entire lifetime. Therefore, the metadata tables for active kernels do not reflect allocations or deallocations that occur during their execution, leaving applications vulnerable to false positives and temporal memory safety errors. Both cuCatch and GPUShield preserve the CUDA ABI, an essential consideration for CPU host interoperability. Thus, *there is a need for an efficient base and bounds scheme that can scale to an arbitrary number of allocations and handle explicit and implicit data transfers without sacrificing temporal safety.*

Table 2: Workload information including instruction counts, number of allocations and memory footprint.

| Workload Name | #Instr. (M) | #Allocations | Footprint (MB) | Workload Name | #Instr. (M) | #Allocations | Footprint (MB) |
|---|---|---|---|---|---|---|---|
| amber18_1 | 283 | 84 | 62 | lammps_4 | 1723 | 49 | 1179 |
| amber18_2 | 60 | 1081 | 3374 | lammps_5 | 150 | 125 | 927 |
| amber18_3 | 64 | 1081 | 3374 | lammps_6 | 1180 | 128 | 1123 |
| amber18_4 | 50 | 92 | 272 | ldpc5Gdecode | 56 | 4 | 2 |
| AMG_1 | 112 | 2 | 473 | namd_1 | 52 | 80 | 2048 |
| AMG_2 | 77 | 1 | 236 | namd_2 | 44 | 1184 | 291 |
| AMG_3 | 1642 | 5 | 957 | Optix1 | 71 | 14 | 1060 |
| AMG_4 | 95 | 5 | 1181 | Optix2 | 94 | 14 | 1060 |
| fun3d | 175 | 6 | 634 | Optix3 | 4469 | 14 | 1060 |
| Laghos_1 | 621 | 6 | 191 | relion_1 | 821 | 22 | 1288 |
| Laghos_2 | 406 | 6 | 191 | RTM_1 | 1024 | 7 | 1675 |
| lammps_1 | 659 | 43 | 530 | RTM_2 | 340 | 9 | 789 |
| lammps_2 | 4905 | 45 | 530 | RTM_3 | 1632 | 13 | 1858 |
| lammps_3 | 3166 | 49 | 1179 | RTM_4 | 2347 | 12 | 1420 |

## 3 GPU Workload Characterization

In order to better appreciate the GPUArmor system design, we first present several metrics of GPU workloads that pertain to memory safety. This section makes several observations regarding the metrics that motivate our eventual solution.

### 3.1 Workloads

We consider 28 CUDA kernels that industry product groups use to guide architecture design. The kernels come from real-world GPU applications, which we simulate using realistic inputs. Our workloads cover different workload segments, from scientific computing (namd, amber18, AMG, FUN3D, Laghos, lammps, Relion, and RTM), to commercial (5G decoding), and visualization (Optix [27]). Section 8 summarizes the profiling and simulation methodology. Table 2 shows each workload's total number of dynamic instructions, number of allocations, and memory footprint.

GPUArmor: A Hardware-Software Co-design for Efficient and Scalable Memory Safety on GPUs

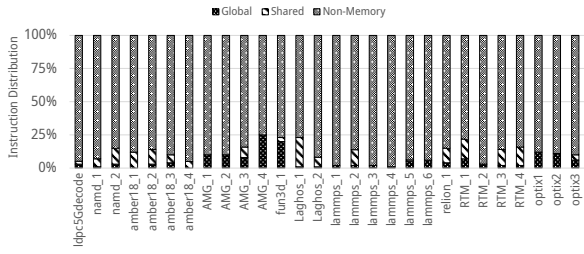

Figure 2: Dynamic instruction distribution showing global, shared and non-memory instructions.

## 3.2 Observations

**Instruction behavior.** Figure 2 shows the dynamic distribution of `global` and `shared` memory operations with respect to all other operations. On average, global memory accesses account for 6% of executed instructions. Compared to typical CPU workloads, this percentage is small [31]. This observation is primarily due to the significant difference in registers available to GPU threads versus a typical CPU, where a GPU compiler can typically avoid spilling and can promote all local variables to registers. In addition, there's an inherent overhead in managing thread-level parallelism.

The CUDA Programming Guide recommends using `shared` memory for frequently touched data where possible, which offers lower latency accesses and a higher bandwidth connection than `global` memory does [23]. We see that the developers of our workloads offload, on average, half of all memory operations to shared memory. Some workloads, such as Laghos make significant use of shared memory, but on average shared memory accounts for roughly 6% of all dynamically executed instructions.

Finally, we note that local memory usage is rare. Figure 2 does not include a category for local memory because none of our workloads has local memory operations. Recall that local memory is used for thread-private data, such as for register spills and local variables, including stack-allocated arrays. Spilling is rare on GPUs because CUDA compilers aggressively inline functions (avoiding the spill/-fill regions ABI calling conventions incur) and the compiler also takes advantage of the massive register files in GPUs. Local array allocations are of limited usefulness because stack allocations cannot be shared with other threads and because the CUDA programming model restricts per-thread stack size (16MB maximum).

**Observation 1**: Global and Shared memory accesses are less frequent in GPU workloads than in CPU workloads, and stack-allocated arrays are practically unused in the GPU paradigm.

**Ramification**: GPU global and shared memory protection schemes may be able to tolerate significant per-memory-operation overheads without commensurately affecting kernel-level overheads. Stack canaries or expensive software-only checking [36] may suffice for protecting stack-allocated arrays.

**Hardware-level behavior.** GPUs are designed to handle highly parallel workloads, which allows them to process many threads simultaneously. When a warp stalls because a memory request misses in the cache, the architecture can switch execution to another warp, effectively hiding memory access latencies.

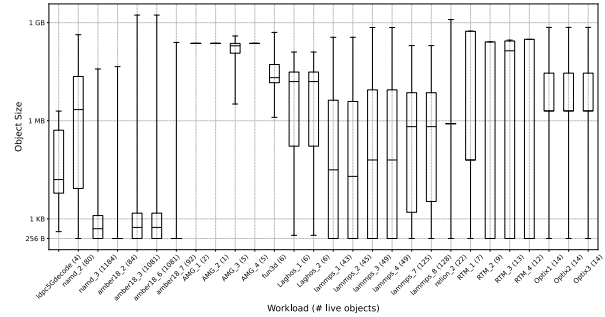

Figure 3: Object size distribution for different workloads.

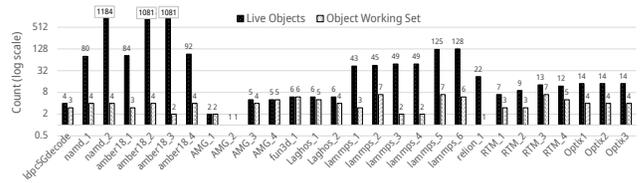

Figure 4: Dynamic object count for different workloads.

**Observation 2**: GPUs are designed to tolerate long-latency operations, especially memory operations.

**Ramification**: GPUs can tolerate the additional latency that memory checking incurs.

GPUs also have limited memory capacity, and Unified Virtual Memory oversubscription is inefficient. NVIDIA's H100 data center part has 80GB of memory, and an A5000 part has 32GB capacity.

**Observation 3**: GPUs have limited memory capacity.

**Ramification**: Metadata structures that scale linearly with memory footprint waste a constraining resource (e.g., cuCatch's 32-bits per 32B [36] or LAK's 8-bits per 16B [40]).

**Memory and allocation behavior.** Allocation sizes impact a memory protection scheme's memory overhead. For instance, if a scheme adds 16B for every allocation to store the allocation's base and bounds, but an application's average allocation size is 16B, then the scheme doubles the application memory footprint. Figure 3 reports that the median allocation size of our workloads varies between 256B and 236MB. For efficiency, CUDA's memory allocator guarantees at least 256B alignment [23], which explains our lower bound. Around 25% of our workloads have a median object size less than 4KB. Further, the maximum object size in our workloads is 1.7GB.

**Observation 4**: GPU allocation sizes tend to be large.

**Ramification**: Small, constant-sized metadata per allocation contributes minor storage overheads.

The maximum number of potentially accessible allocations at any point in a kernel, which we call live allocations, determines the capacity requirements of a memory safety solution's metadata structure. Figure 4 shows that our workload set has three kernels with over one thousand simultaneously live allocations.

Mohamed Tarek Ibn Ziad, Sana Damani, Mark Stephenson, Stephen W. Keckler, and Aamer Jaleel
NVIDIA
{mtarek, sdamani, mstephenson, skeckler, ajaleel}@nvidia.com

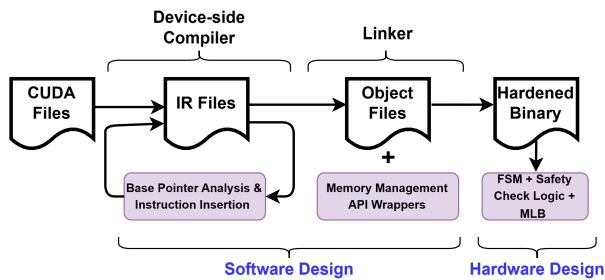

Figure 5: GPUArmor system overview.

**Observation 5**: The maximum number of live allocations in GPU workloads can exceed small, fixed-size hash tables.

**Ramification**: A scalable GPU solution needs to effectively track the metadata of a large number of allocations. Solutions that rely on the upper address bits to index a metadata structure, such as GPUShield [16], cannot reliably account for all potentially reachable allocations for some workloads. With a 57-bit address space, only seven bits can be repurposed as an index, allowing for reliable tracking up to 128 live allocations. Even worse, if temporal safety is added to these schemes (e.g. cuCatch [36]), the available bits drop from seven to three.

On the other hand, while the total number of allocations can be large, the allocation working set tends to be small. We define the allocation working set as the maximal set of simultaneously accessed allocations at all points in a kernel. The working set size represents the number of entries a metadata cache needs to avoid capacity misses. Figure 4 shows that the allocation working set is less than eight for all of our applications.

**Observation 6**: Kernels have small allocation working sets.

**Ramification**: A small per-SM hardware structure with insignificant area cost can effectively cache allocation metadata.

The observations above motivate our GPUArmor design.

## 4 GPUArmor System Overview

To address limitations of prior work, we propose GPUArmor, a hardware-software co-design for efficient and scalable memory safety on GPUs. We first describe the high-level idea of GPUArmor. We then describe its hardware and software details in Section 5 and Section 6, respectively.

### 4.1 Main Idea

Figure 5 provides an overview of our GPUArmor framework. Using CUDA source files as input, GPUArmor performs compile time analysis and instruction insertion (Section 6.2) for the device-side code. The resulting binary is linked against a runtime library that contains wrappers around the memory management APIs (Section 6.1). Finally, the hardened binary and runtime run on a GPU that supports the new instructions (Section 5.1) with memory safety optimizations (Section 5.2).

When an allocation is created, GPUArmor stores the memory safety metadata in a disjoint in-memory structure. To verify the memory safety of an arbitrary pointer at run time, GPUArmor retrieves the memory safety metadata from memory and performs spatial and temporal safety checks. To better understand how GPUArmor works, we consider the metadata life cycle from creation to access and finally deletion, as shown in Figure 1.

### 4.2 Allocation Life-cycle

**Metadata Creation.** At program initialization, GPUArmor reserves, but does not allocate, dedicated address space for the metadata structure. The metadata address space is read-only on the device. In this paper, we experiment with three metadata structures with very different asymptotic behaviors: a linked list, a balanced binary search tree, and a hypothetical (and unrealizable) speed-of-light (SoL) table. The linked list and binary tree store a per-allocation 32-byte node, divided into a 16-byte metadata entry and two 8-byte pointers for next and previous (or child) nodes. The SoL table is an array of 16-byte entries, each of which contains the base and bounds for an allocation. Inspired by cuCatch [36] and GPUShield [16], GPUArmor uses a pointer's value (rather than the pointer's location in memory [19, 24]) as a key to search the metadata structure. This design choice avoids storage overheads and the issue of identifying pointer copy operations (to update per-pointer metadata), which is generally intractable in unsafe languages.

When an application allocates memory (Lines 12 and 13 in Figure 1), GPUArmor stores a 16-byte metadata entry in the metadata structure. The metadata entry consists of an 8-byte *tagged* base address and an 8-byte allocation size. To detect temporal safety violations, GPUArmor tags pointers during the allocation process. That is, in GPUArmor every pointer is a tuple of a randomly generated $t$-bit tag and $v$-bit virtual address. To store the tuple in a 64-bit pointer, which is important for preserving CUDA's existing ABI, the tag size, $t$, depends on the number of available upper pointer bits. It ranges from seven on modern x86_64 architectures with a 57-bit linear virtual address [13] to 16 bits on traditional 48-bit architectures. The remaining $v$ bits are used for storing the virtual base address of the allocation. In addition to tagging the pointer for an allocation, GPUArmor stores the $t$-bit tag in the upper bits of the corresponding metadata entry's base address.

**Metadata Retrieval.** During program execution, when a pointer is used to access memory (Line 8 in Figure 1), GPUArmor validates the safety of the access. To do so, GPUArmor first retrieves the corresponding allocation metadata from the metadata structure using the compiler-identified root pointer (detailed in Section 6). This step might involve multiple memory accesses depending on the metadata structure's layout. A linked list (or binary tree) metadata layout incurs $n$ (or log $n$) 16-byte loads to traverse the linked list (or binary search tree), where $n$ is the position of the allocation metadata entry in the metadata structure. Each load is followed by a comparison to check if the address falls between the allocation base and size of a given node. We elaborate on how this metadata retrieval process is optimized in Section 5.2. For the SoL table, we use an oracle to find a reference to the metadata entry's location in a table, then a single 16-byte load retrieves the corresponding allocation's base, tag, and size information, achieving $O(1)$ performance.

**Metadata Usage for Address Checking.** Once the allocation metadata (base address, size, and tag) is retrieved, we check the

GPUArmor: A Hardware-Software Co-design for Efficient and Scalable Memory Safety on GPUs , ,

memory safety of the corresponding memory access by checking whether the input address lies within the base and size of the allocation (i.e., spatial safety check) and comparing the metadata tag versus the tag from the upper bits of the input address (i.e., temporal safety check). In other words, GPUArmor effectively constructs a fat pointer that holds all the memory safety information without changing the allocation layout or restricting the number of protected allocations to the unused upper pointer bits.

**Metadata Deletion.** When an allocation is deleted (Lines 17 and 18 in Figure 1), GPUArmor invalidates the metadata entry associated with this allocation. This way GPUArmor captures dangling pointer accesses to this region.

## 5 Hardware Design

In this section, we describe the GPUArmor instruction set extensions and the microarchitecture design.

### 5.1 Instruction Set Extensions

To reduce the performance overheads of software-based memory safety, GPUArmor adds three new instructions to the GPU ISA as shown in Table 3. We do not use special instructions for populating the metadata, opting instead to use software-based routines within our runtime wrappers because populating the metadata is an uncommon operation that is executed infrequently (i.e., upon (de)allocation) than operating on the metadata.

**Rd = LOADMETA [Ra].** This instruction takes as input a 64-bit compiler-identified root pointer (Ra) and returns a 64-bit pointer to the metadata location, (Rd) as output. Upon executing LOADMETA, the hardware traverses the metadata structure to find the metadata location associated with the allocation pointed-to by Ra. It then stores the allocation metadata, mdata (i.e., 64-bit base address with tag and 64-bit size) and its 64-bit memory location into a hardware structure dubbed the Metadata Lookaside Buffer or MLB (Section 5.2). LOADMETA does not return the 128-bit metadata entry as output for two reasons: (1) to reduce register pressure and (2) to avoid storing a stale tag which might prevent identification of use-after-free violations in long-running kernels.

To avoid accidentally fetching the mdata of an unrelated allocation, this instruction verifies that (1) the difference between mdata.base and Ra is smaller than mdata.size and (2) Ra's tag matches the mdata.tag. The instruction returns zero otherwise. Similarly, the instruction returns zero if the mdata does not exist without raising an exception to avoid false positives which may happen if a corrupted pointer is computed but never accessed. Note that shared memory base and bounds information is statically determined by the compiler and requires no LOADMETA instruction.

**Rd = MEMCHECK.G Ra, Rb.** This instruction takes as input a 64-bit memory address (Ra) and 64-bit metadata address (Rb), performs bounds and tag checking, and returns an untagged 64-bit memory address (Rd). Upon executing MEMCHECK.G, the hardware uses Rb to fetch the allocation metadata (mdata.base, mdata.size and mdata.tag) either from the MLB if available or from the metadata location in memory (stored in Rb). It then performs bounds check by computing the difference between Ra and mdata.base and comparing it to mdata.size. Next, it compares the upper 7 (or 16 on 48-bit systems) bits of Ra with the fetched mdata.tag to make sure there is no tag mismatch. An exception is raised if (1) the memory access is not within legitimate bounds or (2) there is a tag mismatch.

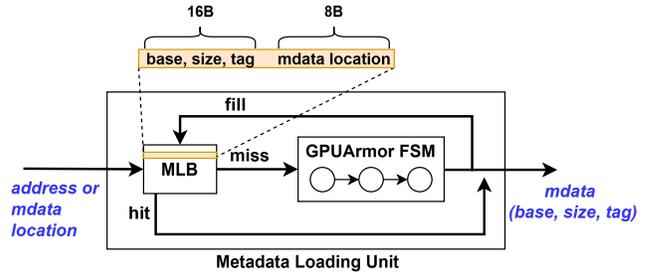

Figure 6: The Metadata Loading Unit (MLU) block diagram.

**MEMCHECK.S Ra, Rb, Rc.** This instruction takes as input a 32-bit shared memory address (Ra), a 32-bit shared memory allocation base address (Rb), and a 32-bit shared memory allocation size, and performs bounds. Since there is no tag for shared memory pointers, the instruction returns nothing. Upon executing MEMCHECK.S, the hardware performs bounds check by computing the difference between the Rb and Ra and comparing it to Rc. An exception is raised if the memory access is not within legitimate bounds.

### 5.2 Microarchitecture Design

When retrieving the memory safety metadata, GPUArmor can incur multiple serial memory accesses, depending on the metadata structure layout (e.g., binary tree or linked list). We implement these memory accesses using a Metadata Loading Unit (MLU) in addition to hardware logic for performing the safety checks.

**Metadata Loading Unit.** The MLU is responsible for retrieving the safety metadata (base address, size, and tag) of the allocation pointed-to by a given memory address. The MLU is co-located with the load-store unit (LSU) in the GPU memory input-output (MIO) block. In the baseline, the LSU coalesces memory accesses across a warp, sends a request to the L1 cache, and upon response writes the data to the corresponding registers. With GPUarmor, the LSU coalesces the LOADMETA addresses and sends it to the MLU which in-turn communicates with the L1 cache. Conceptually, the MLU sits between the LSU and the L1 cache acting as a pass-through for conventional load/store instructions. The MLU traverses the metadata structure to find the location of the metadata entry (mdata.location) associated with the coalesced memory address and returns it to the LSU which writes it to registers. To perform this operation, the MLU implements a finite state machine (FSM), which searches the metadata structure (see Section 4.2). The metadata retrieval occurs in the virtual address space, and the FSM holds the base virtual address for the metadata structure (e.g, head of the link list). The mdata.location is then used by the MEMCHECK instruction to fetch the safety metadata from the L1 cache (or MLB as described below).

**Metadata Lookaside Buffer.** To accelerate metadata retrieval, the MLU contains a Metadata Lookaside Buffer (MLB) which holds the metadata for recently accessed allocations and their locations (Figure 6). Each MLB entry consists of the 16B mdata (base, size, and tag) for an allocation in addition to the 8B virtual address of the metadata entry mdata.location. For LOADMETA instructions, an MLB

Mohamed Tarek Ibn Ziad, Sana Damani, Mark Stephenson, Stephen W. Keckler, and Aamer Jaleel
NVIDIA
{mtarek, sdamani, mstephenson, skeckler, ajaleel}@nvidia.com

Table 3: GPUArmor ISA Extensions.

| Opcode | Inputs | Outputs | Functionality |
| --- | --- | --- | --- |
| LOADMETA | Ra: 64-bit compiler-identified root pointer | Rd: 64-bit pointer to metadata | Traverses the metadata structure using the root pointer and returns the address of the corresponding allocation metadata. Populates the MLB. |
| MEMCHECK.G | Ra: 64-bit global address to-be-checked<br>Rb: 64-bit pointer to metadata | Rd: 64-bit address with tag cleared | Fetches base, tag, and bounds information. Performs global spatial and temporal safety checks and clears the upper address tag bits. |
| MEMCHECK.S | Ra: 32-bit shared address to-be-checked<br>Rb: 32-bit shared allocation base address<br>Rc: 32-bit shared allocation size | - | Performs spatial memory safety checks on shared memory address. |

lookup finds the MLB entry whose range (i.e., [base, base+size]) covers the lookup address, stored in Ra. On a hit, the mdata.location of the matching entry is returned to the LSU. On a miss, the MLU uses the FSM to identify the mdata memory address and sends a request to the cache hierarchy. For MEMCHECK instructions, if an MLB lookup finds the MLB entry whose mdata.location field matches the Rb value, the data is returned to the LSU. On a miss, the MLU uses the pointer to metadata operand (Rb) to load the memory safety metadata from the cache hierarchy. We use a small per-SM 8-entry MLB to reduce power and meet timing requirements. MLB entries are invalidated on deallocations.

**Memory Safety Check Logic.** We implement the logic for performing the spatial and temporal memory safety checks described by the MEMCHECK instruction as an extension to the SM functional unit. The operands of the MEMCHECK instruction are read and operated on as summarized in Section 5.1. If a memory safety error (out-of-bounds access or tag mismatch) is detected, a device-side exception is raised, which can then be captured by the host-side application code. Users can choose to implement an exception handler that either terminates the application or reports the error and resumes execution.

**Binary Compatibility.** To maintain compatibility with code that is non-GPUArmor aware (e.g., accesses to local memory regions in third-party libraries), we leave the 0x0 value unused while assigning the random tags. We only perform the metadata retrieval and safety checks for memory addresses with non-zero tags. Afterwards, our GPU hardware masks off the tag bits before sending the data request to the memory hierarchy similarly to existing CPU features, such as ARM's Top Byte Ignore (TBI), Intel's Linear Address Masking (LAM), and AMD's Upper Address Ignore (UAI).

## 6 Software Design

Here, we describe our changes to the software runtime and compiler in order to enable GPUArmor.

### 6.1 Dynamic Memory Management.

We implement the GPUArmor runtime as wrappers around CUDA's global memory management APIs (e.g., cudaMalloc and cudaFree). When a program allocates memory, we capture the size (s) and base address (p) of the allocation, and also generate a 7- (or 16)-bit tag. We store the size, base address and tag as an entry in the metadata structure, as discussed in Section 4.2. Finally, we add the tag to the upper bits of p and return the tagged pointer p to the program. When a program deletes an allocation, we leave the corresponding entry in the metadata structure as is (base address and size) and set the tag field to a special reserved tag (e.g., 0x7F) to indicate that this memory region has been previously allocated but is currently freed. In this way, dangling pointer accesses to this area (also known as immediate use-after-free) will fail the memory safety check due to a tag mismatch between the reserved tag and the dangling pointer's upper bits.[1] Hence, GPUArmor requires no changes to the memory allocator internals nor does it pose any implementation restrictions. Since the number of allocations and frees is limited, the runtime overhead of metadata management is negligible.

### 6.2 Compiler Support

We implement an instrumentation pass in the GPU-compute compiler back-end to insert the new memory safety instructions: LOADMETA and MEMCHECK described in Section 5.1. The compiler inserts a MEMCHECK instruction before every global or shared memory access (e.g., the global memory read in Line 8 of Figure 1) and a LOADMETA instruction per object being accessed, instead of one per memory access. We define a *root pointer* as the initial pointer from which all other pointers to the object are derived after pointer arithmetic. We then limit LOADMETA insertion to only fetch metadata once for each root pointer. This approach has two advantages: first, we limit the number of memory accesses to fetch metadata; second, we improve the detection of out-of-bounds memory accesses when pointer arithmetic causes an overflow into an adjacent allocation.

**Root Pointer Analysis.** To identify root pointers, we adopt the intra-procedural compiler analysis called the *root pointer analysis* from cuCatch [36]. Our implementation of root pointer analysis performs a recursive search through reaching definitions to find all potential root pointers of an address being accessed. During this backward search, if we find pointer arithmetic, we keep searching until we find a definition that may be a root pointer. For each such reaching root pointer definition, we insert a corresponding LOADMETA instruction, as demonstrated in Figure 1.

For the memory access instruction, R0 = LDG[tmp2] in BB3, our analysis scans backwards through reaching definitions of tmp2 to identify kernel parameters, buf1 and buf2 as the reaching root pointers of tmp2. Next, we insert LOADMETA instructions for both buf1 and buf2 in BB0 to fetch the pointers to the metadata for buf1 and buf2 respectively and propagate the metadata pointers using MOV instructions in BB1 and BB2. Finally, we insert a MEMCHECK instruction before the LDG instruction to perform the memory safety checks using the metadata pointer md_a. Figure 7 shows the total dynamic instruction bloat due to adding our new instructions.

---

[1] If memory (de)allocations exceed a threshold, deleted metadata entries can be reclaimed, though this was not observed in the evaluated workloads.

GPUArmor: A Hardware-Software Co-design for Efficient and Scalable Memory Safety on GPUs

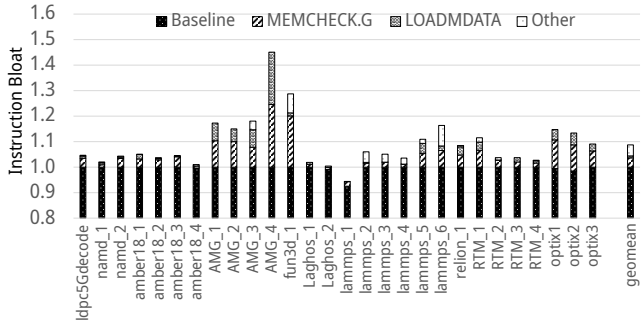

Figure 7: Dynamic instruction bloat of our instrumentation.

Note that due to the limitations of static intra-procedural compiler analysis, the root pointer identified by our analysis may or may not be the actual base address of the object, and instead may represent some other intermediate address within the object. Crucially, the CUDA compiler aggressively inlines device functions, reducing the need for inter-procedural analysis.

**Handling Different GPU Memory Spaces.** While our compiler is able to instrument memory accesses targeting different memory spaces: global, shared, and local, we note that only global memory accesses require a corresponding LOADMETA instruction to fetch the object base and size. This is because the base and bounds information of local and shared memory allocations are known at compile time and are propagated accordingly to the MEMCHECK instructions using a similar compiler analysis.

Finally, we note that our run-time overheads represent an upper bound that may further be reduced with compile-time optimizations proposed by prior work, including bounds check coalescing [36] and statically identifying memory-safe accesses [16].

**Hardware-Only Protection Mode.** In the absence of compiler support (e.g., libraries with no source code availability), GPUArmor runtime and hardware support can still be used for providing reduced memory safety guarantees. We do so by using a mode called GPUArmor-HWOnly, in which we trigger the functionality of LOADMETA and MEMCHECK with every memory access instruction. In this design, the MLU is no longer in pass-through mode, instead each memory instruction fetches the allocation metadata, namely allocation base, size, and tag, and performs the memory safety checks. This approach increases the request rate of memory instructions by fetching the data associated with the memory instruction and also fetching the metadata. We faithfully model this and observe that high hit-rates in the MLB ensure the request rate is similar to baseline. With this metadata, we can deterministically detect spatial (and temporal) safety violations from one allocation to a non-allocated (or recently deleted) allocation and probabilistically detect violations from one allocation to other live allocations in third-party uninstrumented libraries that lack compiler support. While GPUArmor-HWOnly provides similar error detection guarantees as commercialized memory tagging [1, 25], GPUArmor-HWOnly is more memory efficient. A 32KB allocation consumes a 1KB metadata storage with ARM's MTE (4-bit tags per 16B granularity) while consuming 16B with GPUArmor, which is ≈ 64× lower memory cost. Memory savings are higher for large allocations.

Table 4: Memory Safety Error Detection Rates for GPU-based Hardware-assisted Solutions.

| | GPUArmor | GPUShield [16] | GPUArmor-HWOnly | LAK [40] & IMT [33] |
|---|---|---|---|---|
| Adjacent OOB | 100% | 100% | 100% | 100% |
| Non-adj. Intra-scope OOB | 100% | **99.2%** | **99.2%** | **99.2%** |
| Non-adj. Inter-scope OOB | 99.2% | 99.2% | 99.2% | 99.2% |
| Sub-object OOB | **0%** | **0%** | **0%** | **0%** |
| Immediate Use-after-free | 100% | **0%** | 100% | 100% |
| Delayed Use-after-free | 99.2% | **0%** | 99.2% | 99.2% |

## 7 Security Analysis
### 7.1 Threat Model

Due to its low run time and storage costs, we envision GPUArmor to be used during both: pre-deployment (for testing and verification of CUDA applications) and post-deployment (for catching bugs that escape testing) stages. To this end, we consider a threat model that is consistent with prior hardware-based memory safety work on GPUs [16, 33, 40]. Specifically, we assume that the device-side GPU application suffers from one (or more) spatial or temporal memory safety errors that can be abused to illegally read from and write to arbitrary memory addresses. Also, we consider the GPU hardware components to be reliable, and thus side- and covert-channel GPU attacks [9, 20, 30, 41] are out of scope. Finally, our current prototype focuses on protecting device-side code. Existing CPU tools [29] can be used to detect errors in host-side code.

### 7.2 Security Results

Table 4 compares the memory safety error detection rates of GPUArmor against GPU-based hardware-assisted solutions. Red entries are particularly concerning. For consistency, we assume that all schemes have the same number of unused upper pointer bits, $t = 7$ bits on 57-bit architectures.

**GPUArmor Coverage.** As stated in Section 6.2, GPUArmor's compiler analysis can slice through pointer arithmetic operations (which might potentially go out of bounds) and resolve the base address of the pointed-to object (or at least an intermediate non-base address that does not match the input pointer). Thus, it can achieve 100% detection rates for non-adjacent intra-scope[2] OOB errors. Otherwise, GPUArmor relies on memory tagging and hence provides the same probabilistic guarantees as GPUArmor-HWOnly (99.2% detection rate) for catching (1) inter-scope OOB errors that exceed the intra-procedural capabilities of our compiler analysis and (2) delayed UAF errors whose dangling pointers access the deleted memory after being assigned to a new object.[3]

To measure the compiler analysis coverage, Figure 8 shows the percentage of MEMCHECK.G instructions whose LOADMETA is

---
[2]Scope here refers to the view of our intra-procedural root pointer analysis.
[3]Other temporal safety solutions, such as quarantining [29], garbage collection [2], or pointer nullification [37] can be used to deterministically detect delayed UAF errors. While integrating GPUArmor with any of the above schemes is feasible, comparing the different methodologies for addressing temporal memory safety threats on GPUs is beyond the scope of this work.

Mohamed Tarek Ibn Ziad, Sana Damani, Mark Stephenson, Stephen W. Keckler, and Aamer Jaleel
NVIDIA
{mtarek, sdamani, mstephenson, skeckler, ajaleel}@nvidia.com

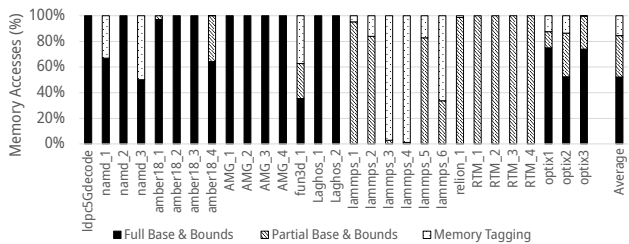

Figure 8: Root pointer analysis coverage for GPUArmor.

using an input operand (i.e., root) that is (1) a true object base address (full base-and-bounds detection), (2) a non-base address that does not match the to-be-checked memory address (partial base-and-bounds detection), or (3) a non-base address that matches the to-be-checked memory address (memory tagging detection). The difference between the first two categories is that in the first (full) category the compiler correctly identifies the true base address of the pointed-to object whereas in the second (partial) category the compiler fails to identify the object base address yet it manages to slice through pointer arithmetic instructions. For the third category, the compiler analysis uses the given memory address itself to load the metadata, which is similar to the GPUArmor-HWOnly functionality. On average, our analysis provides 100% full base and bounds detection rate for 55.5% of the MEMCHECK.Gs, partial detection ($99.2\% < detection < 100\%$) for 36% MEMCHECK.Gs, and memory tagging detection (99.2%) for the remaining 8.5% MEMCHECK.Gs.

**GPUArmor-HWOnly Coverage.** Like existing memory tagging schemes on GPUs [33, 40] and CPUs [1, 25], GPUArmor-HWOnly provides probabilistic memory safety guarantees that depend on the number of unique valid tags, with a detection rate of $100\% - \frac{100\%}{\#\text{Tags}}$. As we assume 57-bit architectures (with the 0x00 and 0x7F tags being reserved for binary compatibility and decorating deleted memory, respectively), GPUArmor has an average detection rate of $100\% - \frac{100\%}{126} = 99.2\%$. The only exceptions are the (1) adjacent OOB errors, which can be 100% detected if different tags are assigned to adjacent objects in memory (e.g., The Android Scudo allocator [17]) and (2) immediate UAF errors, whose dangling pointers access the deleted memory before being assigned to a new object as deleted memory receives a unique tag value of 0x7F. Unlike carve-out based memory tagging schemes [1, 40], GPUArmor-HWOnly can scale its probabilistic guarantees (if more upper address bits are unused) without increasing the storage costs.

**Sub-object OOB Errors.** As shown in Table 4, existing GPU schemes are incapable of detecting sub-object OOB errors that occur between two fields within the same struct. These errors account for around 1% of observed errors on CPUs [14]. Addressing them requires source code instrumentation to either (1) narrow the object bounds whenever an inner buffer field is accessed [19] or (2) promote the inner buffer fields within an object into standalone objects [35]. Integrating these options with GPUArmor is left for future work.

**Concurrent Use-after-free.** Finally, a temporal safety threat that impacts memory safety techniques that load safety metadata into registers (e.g., GPUArmor and cuCatch [36]) is the concurrent UAF, where an object is deleted by one thread while being concurrently used by another thread, leaving the mdata.tag bits stale in the second thread's registers. As this error is relevant for GPU code with thousands of threads/kernels executing in parallel, our prototype always fetches the mdata.tag from the metadata structure entry as part of executing the MEMCHECK.G instruction to avoid using a stale mdata.tag for temporal safety checks.

## 8 Experimental Methodology

**Compilation.** We build our workloads using the nvcc compiler. For evaluating GPUArmor, we modify the backend ptxas compiler to perform the root pointer analysis and emit unique opcodes for the LOADMETA and MEMCHECK.G instructions in the compiled workloads while maintaining the correct register dependency with the corresponding memory instructions. We refer to binaries with metadata instructions as hardened binaries. For evaluating GPUArmor-HWOnly, we use vanilla binaries as the hardware-based memory tagging functionality does not require program modifications. In both configurations, we use the CUDA runtime library with wrappers around memory management APIs for populating the GPUArmor metadata at runtime, as described in Section 6.1.

**Tracing and Simulation.** We collect workload traces for the vanilla and hardened binaries on an NVIDIA GA100 GPU and simulate a GA100 using the NVIDIA Architectural Simulator (NVAS) [39]. The minimum dynamic instruction count per trace is 40 million. We implement GPUArmor, as described in Section 5 and Section 6, respectively. The LOADMETA latency depends on the memory accesses issued by the MLU to identify the metadata location whereas the MEMCHECK latency is bounded by a single memory load to read the metadata from its pre-identified location followed by a fixed three-cycle latency to perform the safety checks. We do not include the performance overheads of the memory management wrappers as they are typically negligible compared to the execution time of the memory management APIs themselves. For GPUArmor we report performance relative to the baseline hardened binary where the memory safety instructions are treated as NOPs. The performance overhead of the hardened binaries relative to the vanilla binaries was negligible (less than 5%). For GPUArmor-HWOnly, we report performance relative to the baseline vanilla binaries.

## 9 Evaluation

In this section, we answer the following research questions:
1. What are the performance overheads of GPUArmor with different metadata structure layouts?
2. How do the hardware optimizations, namely the MLB, improve the GPUArmor performance results?
3. What are the benefits and overheads of GPUArmor in the absence of recompilation?
4. What are the area and energy costs of GPUArmor?
5. How does GPUArmor compare to recent GPU proposals?

### 9.1 GPUArmor Performance Results

**Sensitivity to Metadata Structure.** Figure 9 illustrates the performance sensitivity of GPUArmor to different metadata structures: linked list, binary search tree, and the SoL Table. We observe that workloads with high object counts (e.g., *namd*, *amber*18, *lammps*) incur high performance overheads when using a linked list. On average, a linked list data structure incurs overhead of 2× (up to 22×) due to its O(N) traversal complexity in terms of memory loads. On the other hand, a more efficient data structure like a binary search tree reduces the performance overheads to an average 10% (up to 80%) because of its O(log N) traversal complexity in terms of memory loads. Large allocation count workloads like *lammps* and *relion* still observe high performance overheads due to binary tree traversal latency. Comparatively, the SoL Table incurs a negligible 1.4% performance overhead due to its O(1) traversal complexity. Since the metadata is frequently accessed, we observe high metadata hit-rates (>95%) in the L1 cache, suggesting that overheads are predominantly from L1 cache access latency.

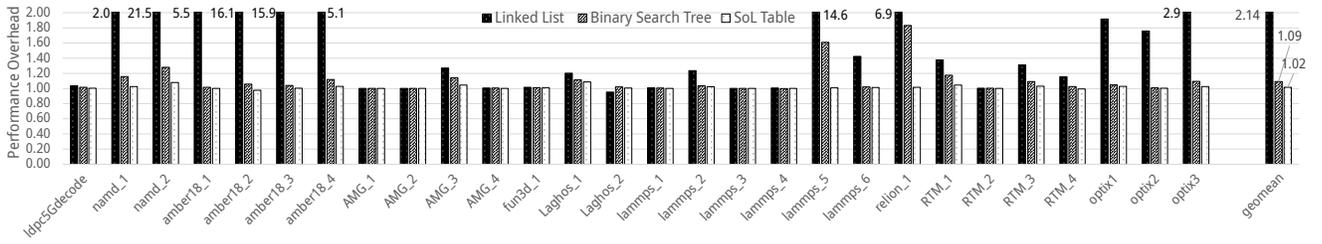

Figure 9: Run time overheads of GPUArmor with different metadata structures.

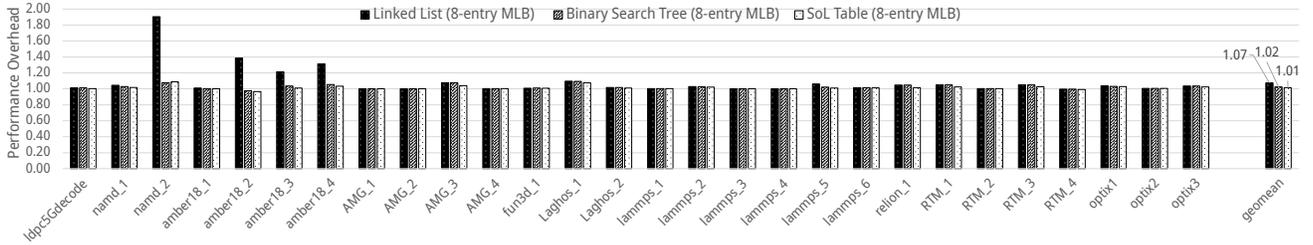

Figure 10: Run time overheads of GPUArmor with different metadata structures (with an 8-entry MLB).

**Sensitivity of Metadata Structure to MLB.** Figure 10 illustrates the performance overhead of the three different data structures in the presence of an 8-entry MLB. Recall, that the MLB enables the LOADMETA instruction direct access to the metadata without traversing the metadata structure. For the majority of our workloads, we observe that an 8-entry MLB hides performance overheads due to metadata traversal. Some workloads like *namd* and *amber*18 incur non-negligible performance overheads with a linked list data structure because of compulsory accesses to the meta data structure (they have over 1000 objects and only 50M instruction run lengths). These results reveal that an adequately sized MLB is high performing and provides independence from the underlying metadata structure (simple or complex).

**Sensitivity to MLB size.** As seen above, an MLB can eliminate performance overheads if it is adequately sized to hold the metadata for the object working set (i.e., the number of objects actively being referenced). Thus, we now study sensitivity of our workloads to number of MLB entries (see Figure 11). The figure shows the minimum, maximum, and geomean performance overheads across all workloads for different MLB sizes. The results show that GPUArmor overheads saturate for an 8-entry MLB and beyond. This correlates with object working set sizes reported in Figure 4.

**Shared Memory Protection Overheads.** Protecting shared memory does not incur additional memory accesses, but compute logic to verify bounds. We modeled shared memory protection by increasing the latency of shared memory accesses to account for the bounds check. Across all our workloads, we observe negligible performance overhead (<0.1%) from the shared memory protection.

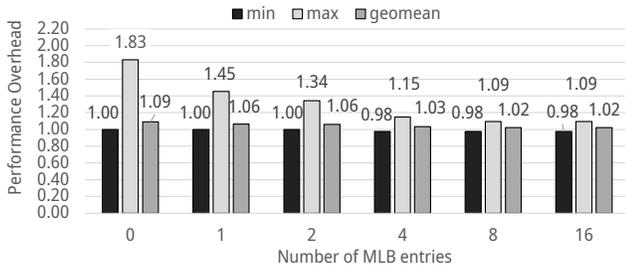

Figure 11: Sensitivity analysis of the MLB.

## 9.2 GPUArmor-HWOnly Results

In the absence of compiler support, GPUArmor provides memory tagging-like error detection coverage, as discussed in Section 6.2. We refer to this mode as GPUArmor-HWOnly in which we operate on vanilla binaries without recompilation. Figure 12 compares the run time overheads of running our workloads with GPUArmor-HWOnly (with an 8-entry MLB) against an ARM's MTE GPU-based implementation (dubbed Lock-and-Key or LAK [40]), which uses a physical memory carve-out for tag storage (8-bit tags per 16B memory region). GPUArmor-HWOnly incurs lower runtime overheads than LAK on average (2.2% versus 10%) with a maximum of

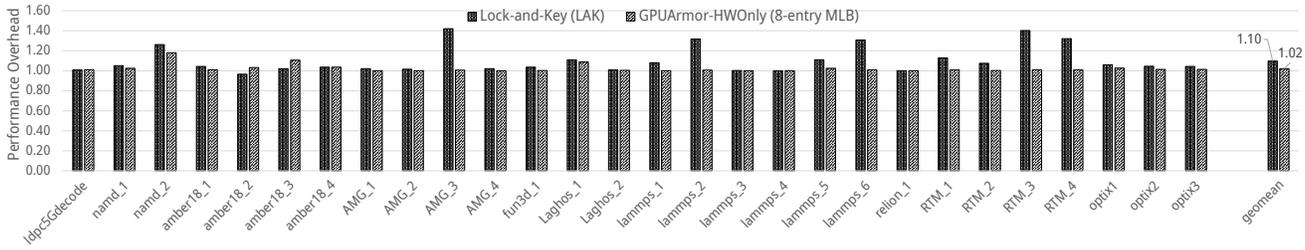

Figure 12: Run time overheads of GPUArmor-HWOnly (with Binary Tree metadata structure and an 8-entry MLB) versus a memory tagging implementation on GPU (LAK [40]).

18.1% versus 42.1%. LAK incurs higher overheads because of additional memory bandwidth required to fetch the metadata (20% in our evaluations). Compared to LAK, GPUArmor-HWOnly requires significantly lower metadata storage overheads for its binary tree (less than 0.001% of total memory footprint) compared to LAK's tag array (with 3.25% memory storage overheads).

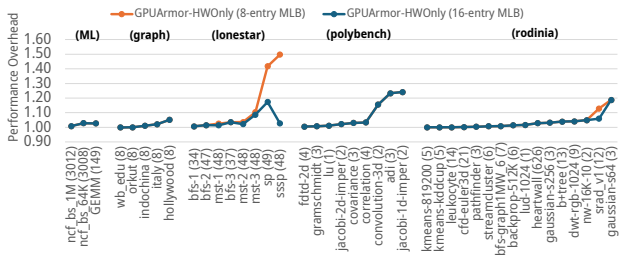

Figure 13: Run time overheads of GPUArmor-HWOnly on conventional workloads with an 8- and 16-entry MLB.

**Results Across Broader Set of Workloads.** We also evaluate GPUArmor-HWonly across a broader set of workloads such as matrix-multiply (implemented in CUTLASS) and NCF from MLPerf, graph processing (page rank), and workloads from the lonestar, polybench, and rodinia benchmark suites.[4] Figure 13 shows the different workloads on the x-axis with the number of live allocations in parenthesis (range between 1 and 3008). Our evaluations show that an 8-entry MLB incurs performance overheads of 6% on average. Workloads like *adi* and *jacobi* incur higher performance overheads due to reduction in effective L1 cache bandwidth to perform the memory safety checks. Performance outliers such as *sp* and *sssp* from lonestar have larger allocation working set sizes requiring a 16-entry MLB. Thus, using a 16-entry MLB incurs average performance overheads less than 4%. These results demonstrate that GPUArmor-HWonly offers a versatile solution for a wide range of workloads, effectively ensuring GPU memory safety with minimal performance impact on average.

### 9.3 Area and Energy Overheads

We now evaluate the area and energy overheads of the 16-entry MLB (each entry stores 24B) and the logic for the FSM within the MLU. This amounts to roughly 384B per SM. Assuming 7nm technology, and an MLB implemented using SRAM, comparators, and priority logic, the MLB is roughly 0.2% the area overhead and 5% the energy per access of the GPU L1 cache. Furthermore, since the MLB is consulted only for global memory instructions (which as Figure 2 shows account for 6% of all instructions executed), its power impact is negligible (<0.0006% of total GPU power). Thus, the MLB incurs negligible overheads.

[4]Due to time constraints, we could not trace these workloads to evaluate GPUArmor.

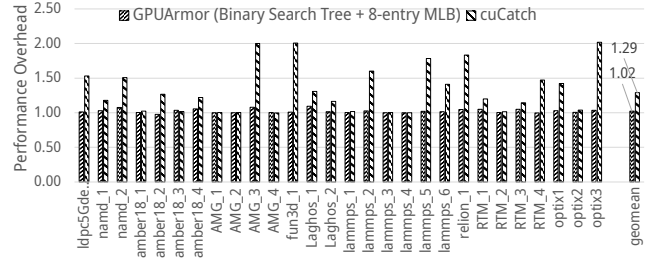

Figure 14: Comparing GPUArmor to cuCatch.

### 9.4 Comparison to Related Work on GPUs

We compare GPUArmor with the recent GPU proposals. For quantitative comparisons, we focus on software-only cuCatch [36] which has identical spatial and temporal safety guarantees but incurs 12.5% metadata storage overhead as compared to the .0005% overheads incurred by GPUArmor (see Table 5). We do not consider clArmor [10] and GMOD [8] since they have lower temporal and spatial safety guarantees and higher runtime overheads on average compared to GPUArmor. We also do not consider GPUShield [16] since it does not (a) provide temporal safety guarantees and (b) cannot scale to workloads with a large number of live allocations.

To ensure a unified platform for performance comparison against cuCatch, we use the cuCatch compiler to compile, run and collect truces of the cuCatch hardened binaries on our GPU simulator. Figure 14 illustrates that cuCatch incurs up to 2X performance overhead, 29% on average. On the other hand, GPUArmor with a binary search tree and 8-entry MLB incurs only a 2% performance overhead. The performance benefits over cuCatch are due to (a) reduction in code bloat and register pressure by avoiding executing the FSM entirely in the compiler and (b) reduction in memory bandwidth and latency by leveraging the MLB. For example, *AMG_3*, GPUArmor reduces the instruction bloat by 2*X* and register pressure by 1.7*X*. The reduction in code bloat and register pressure enables GPUArmor to maintain similar SM occupancy as the baseline. This is a general trend across all workloads where cuCatch has high overheads. Thus, our hardware-software co-designed GPUArmor outperforms the software-only cuCatch with negligible performance and metadata storage overhead.

## 10 Discussion

We now discuss ongoing challenges GPUArmor must consider.

**Supporting Custom Memory Allocators.** GPUArmor currently protects allocations created with the CUDA memory management APIs, such as cudaMalloc. Like other memory safety solutions, GPUArmor will not detect memory safety violations within user-crafted allocations (e.g., subdividing a large allocation into smaller slices via pointer arithmetic tricks), nor can it detect overflows between fields of a structure or class. To support custom memory managers, the GPUArmor runtime could interpose well-defined memory management APIs, such as the do_allocate and do_deallocate hooks in the RAPIDS memory manager [21].

**Supporting Heterogeneous Memory Management (HMM).** HMM simplifies CUDA programming in much the same way that Unified Virtual Memory does [23]. With HMM, GPU kernels can access ordinary host-side allocations made via malloc and new. If HMM is enabled, the number of allocations that *could* (but will likely not) be accessed on the GPU will explode and upset many of the observations we made in Section 3. One potential remedy is to maintain two binary search trees: A CUDA allocation tree, and a CPU allocation tree. The CUDA allocation tree is the tree that GPUArmor already maintains, and includes all CUDA API allocations. The CPU allocation tree contains allocation metadata for allocations that are unlikely destined for the GPU. With two trees, LOADMETA would first search the CUDA allocation tree for a matching allocation, and only when it fails to find an allocation would it traverse the (potentially significantly larger) CPU allocation tree. Given our results with linear, $O(N)$ time complexity traversals, we can extrapolate that the CPU allocation table could contain $2^N$ allocations with reasonable run time overheads. Memory safety aside, it is very important to note that developers already expect significant performance penalties from HMM and UVM.

HMM and UVM also prompt CPU-GPU interoperability concerns. GPUArmor must work for any host system, and NVIDIA GPUs support a diverse set of hosts. GPUArmor cannot tag pointers if the CPU host does not support Top-Byte-Ignore features (e.g., ARM TBI, Intel LAM), because if the host program dereferences a tagged pointer, it will almost certainly crash. Just as with cuCatch, GPUArmor can still bounds check HMM and UVM allocations, but it loses some temporal safety coverage.

**Concurrent Metadata Updates.** We are unaware of CUDA applications that perform allocations meant for an already-running kernel (e.g., a *persistent* kernel that consults a worklist of new allocations). However, the CUDA programming model does not prohibit this behavior, which raises concerns about the thread safety of any associated metadata structure. For example, assume that a call to cudaFree executes concurrently with a kernel. We cannot simply remove the allocation's metadata entry from our metadata structure because thousands of threads might be traversing the structure at any given moment. HMM further complicates this matter because a kernel could access any of the allocations of asynchronously executing CPU threads. While our prototype does not address this issue, we note that there are several possible solutions, including using lock-free data structures, paying the penalty of using atomic operations for synchronization and using *persistent data structure* concepts from the functional programming community.

**Co-design.** GPUArmor is an example of hardware-software co-design. In our prototype, we offload important functionality to LOADMETA and MEMCHECK, but as a result our design's functionality is set. Given the inherent latency tolerance of our instructions, an open question is whether we can support similar functionality by shifting more of the work to software. For instance, instead of supporting cuCatch's memory safety primitives, we could consider creating a programmable software interface for the MLB. Instead of traversing a data structure in a hardware FSM, we could consider running a software routine on misses. Developers might be able to repurpose the hardware in creative ways, perhaps to support associating address ranges with different L2 cache policies [23].

## 11 Conclusion

Recent work shows that GPU applications are vulnerable to memory safety errors. State-of-the-art GPU solutions either trade scalability or memory bloat for performance. In this paper, we challenge traditional wisdom by proposing a scalable, memory efficient, and performant solution for GPUs called GPUArmor. GPUArmor combines a simple compiler analysis with hardware optimizations, leveraging key insights of real-world GPU workloads. We show that GPUArmor achieves base and bounds error detection coverage with a simple binary tree data structure with 2.3% average slowdowns. If recompilation is impossible, GPUArmor achieves memory tagging error detection guarantees with the same hardware and storage cost while maintaining negligible slowdowns (2.2% versus 10%) compared to traditional lock and key implementations.

Mohamed Tarek Ibn Ziad, Sana Damani, Mark Stephenson, Stephen W. Keckler, and Aamer Jaleel
NVIDIA
{mtarek, sdamani, mstephenson, skeckler, ajaleel}@nvidia.com

|  | cuCatch | LAK | Compute Sanitizer/ GPUShield/GPUArmor | GMOD/clArmor | IMT |
|---|---|---|---|---|---|
| **Storage Overhead (%)** | 12.5% | 3.125% | 0.0005% | 0.00026% | 0% |

Table 5: Average metadata storage overhead for our workloads (see Table 2) as a percentage of memory footprint.